\documentclass{article}

\usepackage{PRIMEarxiv}

\usepackage[utf8]{inputenc} % allow utf-8 input
\usepackage[T1]{fontenc}    % use 8-bit T1 fonts
\usepackage{hyperref}       % hyperlinks
\usepackage{url}            % simple URL typesetting
\usepackage{booktabs}       % professional-quality tables
\usepackage{amsfonts}       % blackboard math symbols
\usepackage{nicefrac}       % compact symbols for 1/2, etc.
\usepackage{microtype}      % microtypography
\usepackage{lipsum}
\usepackage{fancyhdr}       % header
\usepackage{graphicx}       % graphics
\graphicspath{{media/}}     % organize your images and other figures under media/ folder

\usepackage{algorithm}
\usepackage{array}
\usepackage[caption=false, font=normalsize, labelfont=sf, textfont=sf]{subfig}

\usepackage{stfloats}
\usepackage{url}
\usepackage{verbatim}

\usepackage{cite}

\usepackage{amsmath, amssymb, amsfonts}
\usepackage{algorithmic}
\usepackage{graphicx}
\usepackage{textcomp}
\usepackage{xcolor}
\usepackage{multirow}
\usepackage{newtxtext}

\usepackage{amsmath}
\usepackage{hyperref}

\begin{document}
    \title{ Representation Learning of Limit Order Book:\\ A Comprehensive Study
    and Benchmarking}

    \author{ Muyao Zhong
    \thanks{M. Zhong is with Harbin Institute of Technology, Harbin 150001, China., M. Zhong is with the Guangdong Key Laboratory of Brain-Inspired Intelligent
    Computation, Department of Computer
    Science and Engineering, Southern University of Science and Technology,
    Shenzhen 518055, China. },
    Yushi Lin
    \thanks{Y. Lin is with the Guangdong Key Laboratory of Brain-Inspired Intelligent
    Computation, Department of Computer
    Science and Engineering, Southern University of Science and Technology,
    Shenzhen 518055, China. },
    Peng Yang
    \thanks{P. Yang is also with the Department of Statistics and Data Science, Southern University
    of Science and Technology, Shenzhen 518055, China. Corresponding
    author: P. Yang. \textit{email: yangp@sustech.edu.cn}, P. Yang is with the Guangdong Key Laboratory of Brain-Inspired Intelligent
    Computation, Department of Computer
    Science and Engineering, Southern University of Science and Technology,
    Shenzhen 518055, China. }.
    }

    \maketitle

    \begin{abstract}
        The Limit Order Book (LOB), the mostly fundamental data of the financial market, provides a fine-grained view of market dynamics while poses significant challenges in dealing with the esteemed deep models due to its strong autocorrelation, cross-feature constrains, and feature scale disparity. Existing approaches often tightly couple representation learning with specific downstream tasks in an end-to-end manner, failed to analyze the learned representations individually and explicitly, limiting their reusability and generalization. This paper conducts the first systematic comparative study of LOB representation learning, aiming to identify the effective way of extracting transferable, compact features that capture essential LOB properties. We introduce \textbf{LOBench}, a standardized benchmark with real China A-share market data, offering curated datasets, unified preprocessing, consistent evaluation metrics, and strong baselines.
        Extensive experiments validate the sufficiency and necessity of LOB representations for various downstream tasks and highlight their advantages over both the traditional task-specific end-to-end models and the advanced representation learning models for general time series. Our work establishes a reproducible framework and provides clear guidelines for future research. Datasets and code will be publicly available at \textit{https://github.com/financial-simulation-lab/LOBench}.
    \end{abstract}

    \section{Introduction}

    The limit order book (LOB) is a crucial data structure in financial markets,
    functioning as a multivariate time series (MTS) that consolidates all outstanding
    limit orders submitted by traders since the opening of the market to the current time.\cite{abergelLimitOrderBooks2016,foucault2005limit}.
    The LOB accumulates outstanding orders that reflect the collective opinions of market participants about current market conditions. Instead of merely capturing transactions with tick-by-tick order flow data, LOB records the evolving intentions of traders over time, offering a rich perspective on market expectations and dynamics. By integrating both the unfilled orders and its evolution, LOB presents a more comprehensive and informative view of the market, making it an essential resource for studying financial market behavior and expectations \cite{t.preisMultiagentbasedOrderBook2006,ntakarisBenchmarkDatasetMidprice2018a}. 
    The enhanced accessibility of LOB data from electronic trading platforms has significantly spurred research efforts to model financial markets based on this rich information source.\cite{zhengTCNLOBTemporalConvolutional2023}.

    \begin{figure}[ht!]
            \begin{center}
                \includegraphics[width=0.48\textwidth]{
                    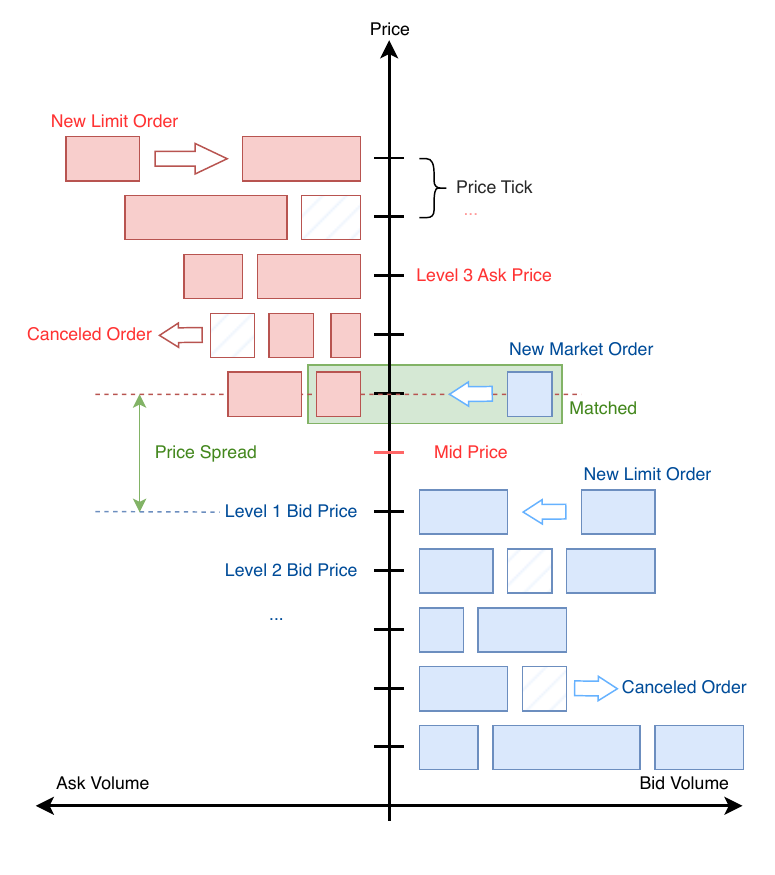
                }
            \end{center}
            \caption{A snapshot of LOB and the trading mechanism. Red blocks represent
            ask (sell) orders, while green blocks denote bid (buy) orders. Both ask
            and sell orders are sorted descending by price. Orders executed at
            the best bid and ask prices are highlighted in blue, while canceled orders
            are masked by slashes. }
            \label{fig:lob_example}
    \end{figure}

    A Limit Order Book (LOB) provides a structured, real-time view of buy and sell orders in electronic markets, built through a continuous matching mechanism that handles three main types of orders: limit orders, which specify price and quantity and remain in the LOB if unmatched; market orders, which execute immediately at the best available price and do not reside in the LOB; and cancellation orders, which remove existing unmatched limit orders.

    The LOB consists of the bid side, which contains unmatched buy orders representing the highest prices buyers are willing to pay, and the ask side, which contains unmatched sell orders representing the lowest prices sellers are willing to accept.

    Each side is further organized by price levels, where each level corresponds
    to a distinct price and the aggregated volume of all the unmatched orders specified with this price. All the aggregated orders at each price level are organized in a queue in terms of the arrival time. Fig.~\ref{fig:lob_example}
    provides an illustrative snapshot of a typical LOB: the horizontal axis represents
    price levels, and the vertical axis indicates the corresponding cumulative
    order volume. Higher levels capture finer-grained market details, although
    accessing such depth generally requires additional data privileges. 

    Within a trading day, these hierarchical bid and ask queues evolve over time as new orders enter the
    market and existing ones are either matched or canceled. This continuous
    update mechanism results in a time series of LOB states, offering insights into
    short-term supply and demand dynamics. Owing to its granular nature, LOB data
    naturally exhibits strong temporal dependencies and cross-feature correlations,
    making it a compelling, yet challenging, research object for advanced
    representation learning techniques.
    
    LOB data can be naturally structured as a three-dimensional tensor, in which 
    the feature dimension (bid price, bid volume, ask price, and ask volume) and level dimension (different price levels) construct a snapshot of the market with 
    price and volume features across multiple levels. By stacking the snapshots along the
    temporal dimension, LOB captures the evolution of market dynamics over time\cite{abergelLimitOrderBooks2016}.
    A LOB snapshot is illustrated in Fig.~\ref{fig:lob_example}, where
    the bid and ask prices and volumes are displayed at different price levels. The
    LOB organizes orders into distinct price levels with a strict hierarchy: for
    bids, orders with higher prices are given priority, while for asks, lower prices
    take precedence. Orders at the same price level are organized as a queue, in terms of their arrival time.

    Formally, LOB data can be formulated as an MTS as follows: 
    \begin{equation}
        \mathbf{S_t}= [\mathbf{L}_{t}, \mathbf{L}_{t+1}, \cdots, \mathbf{L}_{t+T-1}
        ] \in \mathbb{R}^{T \times l \times 4 }.
    \end{equation}
    and each snapshot $\mathbf{L}_{t}$ is represented as: 
    \begin{equation}
        \mathbf{L}_{t}=
        \begin{bmatrix}
            b_{1}^{p}     & b_{1}^{v}     & a_{1}^{p}     & a_{1}^{v}     \\
            b_{2}^{p}     & b_{2}^{v}     & a_{2}^{p}     & a_{2}^{v}     \\
            \vdots        & \vdots        & \vdots        & \vdots        \\
            b_{l}^{p} & b_{l}^{v} & a_{l}^{p} & a_{l}^{v}
        \end{bmatrix}
        \in \mathbb{R}^{l \times 4}.
    \end{equation}
    \noindent For each time step $t$, let us denote:
    \begin{itemize}
        \item $b_{i}^{p}$,$a_{i}^{p}$ as the bid or ask price at the $i$-th
            level;

        \item $b_{i}^{v}$,$a_{i}^{v}$ as the aggregated bid or ask volume of the orders at the $i$-th
            level;

        \item $l$ as the number of price levels in LOB, which is set to
            $5$ for default users in china A-share market, and $10$ or higher for
            paid customers.
    \end{itemize}

    As an MTS, LOBs inherit several common challenges in modeling, including high-dimensional, non-stationary, and noisy. The dimensionality grows with both the number of price levels and the trading horizon length, making direct modeling computationally intensive \cite{lu2018high}. Moreover, order execution mechanisms drive frequent and abrupt changes in the LOB, as new orders dynamically update the book \cite{rocsu2009dynamic}. In addition, strategic behaviors such as frequent cancellations introduce distinctive noise patterns that stem from agent-level decision-making, rather than from conventional measurement errors \cite{eisler2012price}.
    In the literature, a variety of Deep Learning (DL)-based strategies have been developed for generic MTS modeling, including Long Short-Term Memory (LSTM) \cite{tsantekidis2017lstm}, Convolutional Neural Networks (CNNs) \cite{tsantekidis2020cnn2}, Autoencoder \cite{nousi2019machine}, and Transformer-based architectures \cite{wallbridge2020translob,arroyo2024deep}, have demonstrated strong capabilities in automatically extracting intricate patterns from MTS data, establishing new baselines for various prediction and imputation tasks \cite{tangSurveyMachineLearning2022b}.

    % Traditional methods focus on hand-crafted features (e.g., moving averages, price spreads, volatility) \cite{cont2010stochastic}, but they often struggle to capture complex temporal dependencies \cite{kirilenko2017flash,tripathi2020limit}. 

    However, despite these advances, LOB data presents unique challenges  (namely, strong
    autocorrelation\cite{gould2013limit}, cross-feature constrains, and disparity
    in feature scales\cite{li2024simlob}) that differ from generic MTS settings, requiring more 
    specialized modeling approaches\cite{cranmerFrontierSimulationbasedInference2020,bouchaud2018trades,bamfordMultimodalFinancialTimeseries2023a}. 
    Specifically,
    \begin{itemize}
        \item Strong autocorrelation: Let $\mathcal{O}_{t}^{t+1}$ denote the set of orders arriving
            between time frame $t$ and $t+1$, and $\mathcal{C}$ be the Continue
            Double Auction(CDA) mechanism employed by the LOB market. For clarity,
            CDA is a price-time priority order matching system that executes
            orders based on the best available price and their submission times.
            Then the LOB update can be denoted as Eq.\ref{eq:lob_update}.
            This inherent dependence between consecutive snapshots induces non-linear
            temporal autocorrelation, demanding sophisticated temporal modeling to
            capture evolving market dynamics. For a detailed explanation of the LOB construction process, readers are referred to Section~\ref{sec:pre_lob} and the comprehensive treatment provided in \cite{abergelLimitOrderBooks2016}.
            
            \begin{equation}
                \mathbf{L}_{t+1}= \mathcal{C}(\mathbf{L}_{t}, \mathcal{O}_{t}^{t+1}
                ) , 
                \label{eq:lob_update}
            \end{equation}
            
            \begin{equation*}
            \mathcal{O}_{t}^{t+1}=\{{o}_{t+\delta_1}^{1},\cdots,{o}_{t+\delta_j}^{i}
            ,\cdots,{o}_{t+1}^{m}\},
            \end{equation*}
            \begin{equation*}
                0<\delta_{1}<\cdots<\delta_{j}<\cdots<\delta_{m}\leq 1 .
            \end{equation*}

        \item Cross-feature constraints: The LOB enforces strict
            monotonicity among bid price $\{b_{i}^{p}\}$ and ask price 
            $\{a_{i}^{p}\}$:
            \begin{equation}
                b_{1}^{p}> b_{2}^{p}> \cdots > b_{Level}^{p},
            \end{equation}
            \begin{equation}
                a_{1}^{p}< a_{2}^{p}< \cdots < a_{Level}^{p},
            \end{equation}
            \begin{equation}
                b_{1}^{p}<a_{1}^{p}.
            \end{equation}
            These relationships ensure that overlapping or inverted bid-ask
            prices never occur in valid market data. Violating them leads to
            unrealistic LOB representations. Incidentally, it is also worth
            mentioning that all elements in $\mathbf{L}_{t}$ must be strictly
            positive, as prices and volumes must be nonnegative in a valid trading
            environment.

        \item Feature scale disparity: Price and volume can differ by
            several orders of magnitude. For instance, $b_{i}^{p},a_{i}^{p}$ may
            span a few cents to several dollars (or beyond, depending on the
            assets), while $b_{i}^{v},a_{i}^{v}$ typically reflect the number of
            shares or contracts which can be in the hundreds even millions. Such disproportionate
            feature scales complicate normalization and run the risk of biasing modeling
            emphasis on part of dimensions over the others.
    \end{itemize}

   These unique challenges introduce complexities that existing DL methods for general MTS fail to 
    capture effectively. In response, recent research has proposed
    specialized models (e.g., DeepLOB\cite{zhang2019deeplob}, SimLOB\cite{liyuanzhe2024simlob}),
    explicitly designed to align with the structure and dynamic characteristics of LOB data. Empirical evidence consistently shows that such tailored models can help significantly outperform generic models on certain downstream tasks, underscoring the necessity of developing domain-specific solutions to fully leverage the rich information embedded in LOB data.
    In the literature, most LOB modeling efforts have focused on tightly coupling model architectures to specific downstream tasks. This design paradigm, although limiting in terms of flexibility and reusability, has been widely adopted because the close alignment between model structure and task-specific objective often yields immediate and substantial performance gains.
    
    However, task-specific architectures inevitably introduce substantial human costs, as they require repeated design efforts for different downstream tasks. Moreover, the lack of standardized practices in LOB data handling ( e.g., data formatting, normalization, segmentation, and labeling), and the absence of consistent evaluation protocols for downstream tasks, further exacerbates the problem. These inconsistencies not only hinder the reusability and generalization of models, but also make it difficult to conduct fair and meaningful comparisons across different studies.
    
\begin{figure}[!ht]
        \begin{small}
            \begin{center}
                \includegraphics[width=0.45\textwidth]{
                    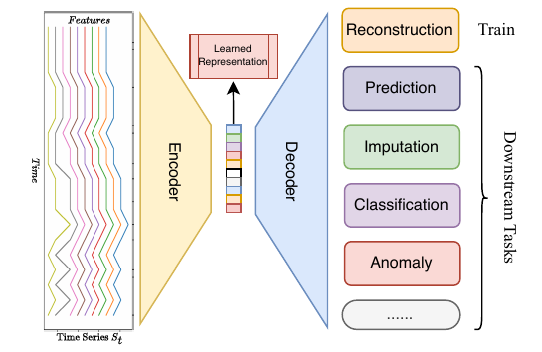
                }
            \end{center}
            \caption{The basic pipeline of representation learning for MTS.}
            \label{fig:pipeline}
        \end{small}
    \end{figure}

    Observing the commonalities across various financial applications (e.g., price trend forecasting\cite{jin2020prediction}, anomaly detection\cite{xu2021anomaly}, and market impact estimation\cite{mertens2022liquidity}) reveals that all of them rely on extracting meaningful abstractions from the raw LOB data. This
    naturally motivates the development of LOB representation learning, a strategy aimed at mapping LOB inputs into compact, informative representations that preserve essential structural and dynamic properties. Formally, given a LOB sequence $\mathbf{S}_t$, representation learning seeks to learn a mapping function $\mathcal{F}$ such that:
    
    \begin{equation}
        \mathcal{R} = \mathcal{F}(\mathbf{S}_t) \in \mathbb{R}^d, d \ll T\times l\times 4,
    \end{equation}
    where $\mathcal{R}$ is the $d$-dimensional representation capturing the salient features of time $t$. Downstream tasks can then be performed based on $\mathcal{R}$ through the generic solver $\mathcal{G}$:
    
    \begin{equation}
        \hat{y}_t = \mathcal{G}(\mathcal{R}),
    \end{equation}
    where $\hat{y}$ denotes the output of the specific downstream task. The basic pipeline of this paradigm is shown in Fig.~\ref{fig:pipeline}. By learning robust and transferable LOB representations, researchers can significantly reduce the burden of bespoke feature engineering \cite{cont2010stochastic} and enable the application of general-purpose time series models across different datasets and market assets. This paradigm offers a direction to balance model performance and development costs, and shift promises to improve reusability and generalization in LOB-based financial modeling.
    
    Building an effective LOB representation learning framework requires addressing several foundational challenges. 
    First, to facilitate fair comparisons across different models and studies, it is essential to establish standardized definitions for LOB data and associated tasks, including data formatting, preprocessing, and evaluation protocols for various downstream applications. 
    Second, to ensure the validity of representation learning itself, one must rigorously evaluate whether the learned representations sufficiently capture the aforementioned distinctive characteristics of LOB data.
    Third, beyond demonstrating that representation learning is technically feasible, it is important to establish its practical necessity, namely, how and why learned representation can offer tangible advantages over task-specific designs.

    In this paper, we undertake a comprehensive study of LOB representation learning
    and introduce a dedicated evaluation framework based on real China A-share market
    data. Our main contributions are summarized as follows:
    \begin{itemize}
        \item We establish the first standardized LOB representation learning benchmark, called \textbf{LOBench}, which provides unified definitions for data formatting, preprocessing, and task-specific evaluation metrics, thereby enabling reproducible, fair and meaningful comparisons across models.
        \item We conduct a systematic investigation into the \textbf{sufficiency} of LOB representations, rigorously analyzing whether different approaches effectively can capture key LOB-specific properties.
        \item We empirically evaluate the \textbf{necessity} of adopting LOB representation learning over traditional task-specific designs, demonstrating its potential for improving model generalization, reducing development costs, and facilitating transferability across different datasets and financial tasks.
    \end{itemize}

    The remainder of this paper is organized as follows. Section II reviews related work on MTS representation learning and surveys on LOB modeling. Section III introduces LOBench, our standardized benchmark designed to enable consistent and reproducible evaluations for LOB representation learning. Section IV describes the experimental setup and presents extensive empirical results addressing the key research questions outlined earlier. Finally, Section V concludes the paper and discusses promising directions for future research.

    \section{Related Work}

    \subsection{Representation Learning }
    Representation learning has emerged as a foundational approach for modeling complex data across domains such as finance\cite{sezerFinancialTimeSeries2020a},
    healthcare\cite{xu2018raim}, and industrial fault diagnosis\cite{zhang2021multiple}.
    By learning compact and informative features from raw inputs, it enables improved performance on variety of downstream tasks, including forecasting\cite{cai2024msgnet,wang2024timemixer},
    classification\cite{middlehurst2024bake,mohammadi2024deep}, regression\cite{mohammadi2024deep,mulla2024times},
    and anomaly detection\cite{choi2021deep}.  
    
    In the context of MTS, representation learning is particularly valuable due to the inherent high-dimensional,
    noisy, and non-stationary \cite{zhangSelfsupervisedLearningTime2024} properties. Early methods (e.g.,  Fourier transform, Wavelet transform, and
    Symbolic Aggregate Approximation) laid important groundwork\cite{wangDeepTimeSeries2024}, but 
     they rely heavily on domain-specific heuristics and  struggle to capture complex temporal structures in high-frequency data\cite{triratUniversalTimeseriesRepresentation2024}.

    To overcome these limitations, DL has emerged as a powerful alternative,
    offering the ability to automatically extract meaningful features directly from raw sequences. Recurrent
    Neural Networks (RNNs), particularly LSTM\cite{niu2020lstm}
    and GRU \cite{pirani2022comparative} architectures, were among the first to gain traction in MTS tasks due to their strong capacity to model temporal dependencies\cite{mao2022review, pirani2022comparative}. CNNs have also been successfully applied, enabling efficient pattern extraction through parallel operations\cite{jin2020prediction}.
    More recently, Transformer-based architectures, originally developed in nature language processing, have been adapted to the time series domain \cite{zerveas2021transformer, xu2021anomaly}, offering scalability and long-range dependency modeling.
    Hybrid models that combine the strengths of RNNs, CNNs, and attention mechanisms further improved performance \cite{livieris2020cnn, khan2021bidirectional,
    xie2020evolving}.

    While recent efforts have increasingly focused on learning general-purpose, transferable representations for time series, many existing MTS methods are still predominantly developed with specific downstream tasks in mind.  As a result, their generalization ability across
    tasks and datasets remains limited. This limitation becomes particularly pronounced in the context of 
    LOB data, which exhibits domain-specific characteristics mentioned earlier.
    
    Existing LOB-specific models have been proposed for various 
    tasks, including price trend prediction \cite{yinDeepLOBTrading2023,zhao2024quantfactor}, market simulation \cite{jainLimitOrderBook2024,wang2025alleviating}, and model calibration \cite{li2024simlob,yang2025calibrating}.
    However, these methods are typically designed in a task-specific, end-to-end fashion, tightly binding the dataset and downstream objective. Consequently, there is no unified framework for explicitly learning or evaluating LOB representations.
    This lack of standardization, both in how LOB data is preprocessed and how representations are assessed, makes it difficult to compare models and limits methodological progress.

    \begin{table*}
    [htbp]
    \centering
    \caption{Comparison of Related Surveys and Research Gap}
    \label{tab:related_survey}
    \setlength{\tabcolsep}{5pt} % 调整列间距（可选）
    \begin{tabular}{|l|c|c|c|c|c|c|c|c|c|c|c|}
        \hline
        \textbf{Survey}                        & \multicolumn{3}{|c|}{\textbf{Tasks}} & \multicolumn{4}{|c|}{\textbf{Models}} & \multicolumn{2}{|c|}{\textbf{Datasets}} \\
        \hline
                                               & \textbf{Prediction}                     & \textbf{Classification}                     & \textbf{Imputation}                      & \textbf{MLP} & \textbf{CNN} & \textbf{RNN}  & \textbf{Transformer} & \textbf{MTS} & \textbf{LOB} \\
        \hline
        Fawaz et al. (2019)\cite{fawaz2019deep}       & \checkmark                           & \checkmark                            &                                        &              & \checkmark   & \checkmark    &                & \checkmark   &              \\
        \hline
        Braei et al. (2020)\cite{braei2020anomaly}    &                                      & \checkmark                            &                                        &              & \checkmark   & \checkmark    &                & \checkmark   &              \\
        \hline
        Lim et al. (2020)\cite{lim2021time}           & \checkmark                           &                                       &                                        & \checkmark   &              & \checkmark    &                & \checkmark   &              \\
        \hline
        Torres et al. (2021)\cite{torres2021deep}     & \checkmark                           &                                       &                                        & \checkmark   & \checkmark   &               &                & \checkmark   &              \\
        \hline
        Wen et al. (2022)\cite{wen2022transformers}   &                                      &                                       &                                        &              &              &               & \checkmark     & \checkmark   &              \\
        \hline
        Tang et al. (2022)\cite{tangSurveyMachineLearning2022b}       &                                      & \checkmark                            &                                        &              & \checkmark   &               &                & \checkmark   &              \\
        \hline
        Shao et al. (2023)\cite{shao2024exploring}    & \checkmark                           & \checkmark                            &                                        &              & \checkmark   & \checkmark    & \checkmark     & \checkmark   &              \\
        \hline
        Meng et al. (2023)\cite{meng2023unsupervised} & \checkmark                           & \checkmark                            &                                        &              & \checkmark   &               & \checkmark     & \checkmark   &              \\
        \hline
        Wang et al. (2024)\cite{wang2024deep}         & \checkmark                           &                                       &                                        &              &              & \checkmark    & \checkmark     & \checkmark   &              \\
        \hline
        Ours                                   & \checkmark                           & \checkmark                            & \checkmark                             & \checkmark   & \checkmark   & \checkmark    & \checkmark     & -            & \checkmark   \\
        \hline
    \end{tabular}
\end{table*}

    \subsection{Related Surveys}
    Over the past decade, numerous surveys have reviewed advances 
    in MTS representation learning from various perspectives. As summarized in Tab.~\ref{tab:related_survey}, early work such as
    Fawaz et al. \cite{fawaz2019deep} and Braei et al. \cite{braei2020anomaly}
    focused primarily on classification and anomaly detection, using RNNs
    and CNNs, while paying little attention to financial domains like LOB. Later surveys by
    Torres et al. \cite{torres2021deep} and Wen et al.
    \cite{wen2022transformers} highlighted the potential of Transformer-based models, yet did not address the cross-feature dependencies that are 
    intrinsic to LOB data. Other efforts, including Tang et al. \cite{tangSurveyMachineLearning2022b}, Shao et al.
    \cite{shao2024exploring}, and Meng et al. \cite{meng2023unsupervised}, examined hybrid architectures and evaluation protocols, but remain focused on general MTS tasks such as forecasting.
    
    Although Wang et al. \cite{wang2024deep} specifically discusses financial applications and introduces GNN-based models, no prior survey or benchmark has systematically addressed LOB representation learning. In particular, there is a lack of standardized datasets, preprocessing procedure, and evaluation metrics tailored to the unique structure of LOB.

    This absence of a dedicated benchmark poses a major obstacle to progress in LOB-related research.
    Without a common foundation for evaluating and comparing representation learning approaches, it is difficult to assess model quality or facilitate reproducible research. To fill this gap, we propose \textbf{LOBench}, a standardized benchmark framework designed to advance the study of LOB representation learning through consistent datasets, metrics, and tasks.

    \section{LOBench Framework}
    LOBench is a standardized benchmark framework
    designed to address the lack of consistency and comparability in LOB representation learning research. 
    It provides a unified foundation for evaluating models across multiple tasks and datasets, enabling reproducible and fair comparisons. Specifically, LOBench offers: 1) curated LOB datasets from the real-world China A-share market, 2) configurable data processing pipelines that standardize normalization, segmentation, labeling procedures, and 3) task-specific evaluation metrics for assessing key downstream applications such as reconstruction, prediction and imputation.

    By formalizing the experimental setting and decoupling data representation from downstream objectives, LOBench facilitates systematic investigation into the sufficiency, necessity, and methodology of LOB representation learning. We aim to provide the research community with a rigorous and extensible platform that supports both benchmarking and innovation, and that lays the groundwork for future advances in this field.

    \subsection{Dataset: China A-share Market}

    While the widespread use of Western financial markets datasets in academic research, such as the widely adopted FI-2010\cite{ntakarisBenchmarkDatasetMidprice2018a} dataset, the China A-share market offers a markedly different yet underexplored environment with substantial research value. Its distinctive institutional settings (e.g., the T+1 non-intraday settlement mechanism, daily price fluctuation limits, and unique trading halts) shape liquidity and volatility in ways that differ from market like the New York Exchange (NYSE) and NASDAQ. Moreover, the dominance of retail investors and policy-driven behaviors introduces dynamics not typically observed in institution-heavy Western markets.

    Despite its scale and complexity, the China A-share market remains significantly underrepresented in publicly available LOB datasets. To our knowledge, there exists no comprehensive,
    well-structured dataset tailored for this market, posing a barrier to reproducible and generalizable research. To address this gap, we construct and release a curated datasets based on real China A-share LOB data, enabling the community to study a broader spectrum of market structures and to develop more versatile representation learning models.
    
    Due to the sensitivity and proprietary nature of the raw order flow data, we are unable to release it
    in its original form. Instead, we provide desensitized LOB
    snapshots sequences derived from the continuous matching process, which preserve the core structural and temporal characteristics of the market. The current dataset focuses on a small but diverse set of representative A-share stock, and we plan to expand its coverage in future releases to further support empirical research on LOB modeling in emerging markets.

    \subsection{Data Preprocessing}
    \label{sec:data_preprocessing}

    \subsubsection{Selecting}
    To construct the dataset for our study, we first obtained the full-year trading
    records for thousands of stocks listed on the Shenzhen Stock Exchange (SZSE)
    in 2019. Following SZSE trading rules, we aggregated each LOB snapshot at 10
    levels with a 3-second sampling frequency, resulting in a high-resolution dataset
    that captures dynamic market conditions throughout the year.

    From this large pool of stocks, we then applied a selection strategy
    inspired by \cite{prata2024lob} to identify five candidates that exhibit
    robust liquidity characteristics. Specifically, we ranked the stocks by
    trading volume, volatility, and average bid-ask spread, ultimately selecting
    Ping An Bank (sz000001), China Vanke (sz000002), Gree Electric Appliances (sz002415),
    Wuliangye (sz000858), and Guangzhou Xiangxue Pharmaceutical (sz300147).
    These companies span multiple sectors (e.g., finance, real estate, consumer electronics,
    and beverages), providing substantial diversity in market behavior and
    volatility. Further details on these stocks are presented in Tab.~\ref{tab:stock_stats}.

    This targeted selection ensures that our benchmark dataset captures representative
    and challenging market conditions for evaluating model performance.
    Additionally, we plan to augment this dataset by gradually including more A-share
    LOB data to enhance its coverage and support a wider range of research inquiries.

    \small

    \begin{table*}[!ht]
        \centering
        \caption{Statistics of the selected China A-share stocks in 2019.}
        \begin{tabular}{|c|c|c|r|r|r|r|r|r|}
        \hline
        \multicolumn{3}{|c|}{Stock Information}& \multicolumn{4}{c|}{Price (RMB)} & \multicolumn{2}{c|}{Volume Mean} \\
        \hline
        Company Name &Ticker  & Sector & Mean & Std & Max & Min & Bid Side & Ask Side \\
        \hline
        \textbf{PingAn Bank} & sz000001 & Finance & 13.83 & 1.91 & 18.29 & 9.10 & 1864194 & 1964112 \\
        \hline
        \textbf{China Vanke} & sz000002 & Real Estate & 27.88 & 1.72 & 33.70 & 23.68 & 483086 & 542093 \\
        \hline
        \textbf{Wuliangye} & sz000858 & Beverage & 108.06 & 26.52 & 154.00 & 46.06 & 63215 & 61627 \\
        \hline
        \textbf{Gree Electric Appliances} & sz002415 & Appliances & 31.01 & 3.16 & 38.61 & 22.77 & 303046 & 248262 \\
        \hline
        \textbf{Guangzhou Xiangxue } & sz300147 & Pharmaceutical & 7.00 & 0.83 & 10.10 & 3.71 & 496117 & 304893 \\
        \hline
        \end{tabular}
        \label{tab:stock_stats}
    \end{table*}
    \normalsize

    \subsubsection{Temporal Alignment and Cleaning}
    To preprocess the LOB, we construct snapshots at a 3-second frequency from raw
    order flow records, and each snapshot includes bid/ask prices/volumes for
    $l = 10$. Furthermore, for time periods where no trades occurred (and thus
    no LOB snapshots were recorded in the raw data), we apply forward-filling by
    duplicating the most recent LOB snapshot. This ensures that the resultant
    dataset maintains temporal continuity and completeness. Furthermore, the
    trading rules of SZSE include a centralized auction phase prior to market
    opening (from 9:15 to 9:30) and shortly before the market close (from 14:57 to 15:00). During this phase, the LOB data is often noisy and may violate
    fundamental LOB constraints. For instance, certain price levels may remain
    unfilled, with bid and ask prices recorded as zero, resulting in equal
    prices across multiple levels. To address these inconsistencies, we discard
    the LOB data generated during the centralized auction phase and only consider the continuous auction phase (from 9:30 to 11:30, and from 13:00 to 14:57).

    \subsubsection{Normalizing}
    Normalization is a fundamental step in machine learning pipelines, commonly achieved
    by applying z-score transformations to each feature independently. In a
    typical feature-wise approach (for instance, as done in the FI-2010 dataset),
    each feature $x_{j}$ is standardized using 
    \begin{equation}
        \label{eq:per_feature_normalization}x_{ij}' = \frac{x_{ij}- \mu_{j}}{\sigma_{j}}
        ,
    \end{equation}
    \normalsize where $x_{ij}$ denotes the $i$-th price or volume for feature $j$,
    and $\mu_{j}$ and $\sigma_{j}$ are the mean and standard deviation of that
    feature, respectively.

\begin{figure}[htpb]
        \centering
        \includegraphics[width=0.49\textwidth]{
            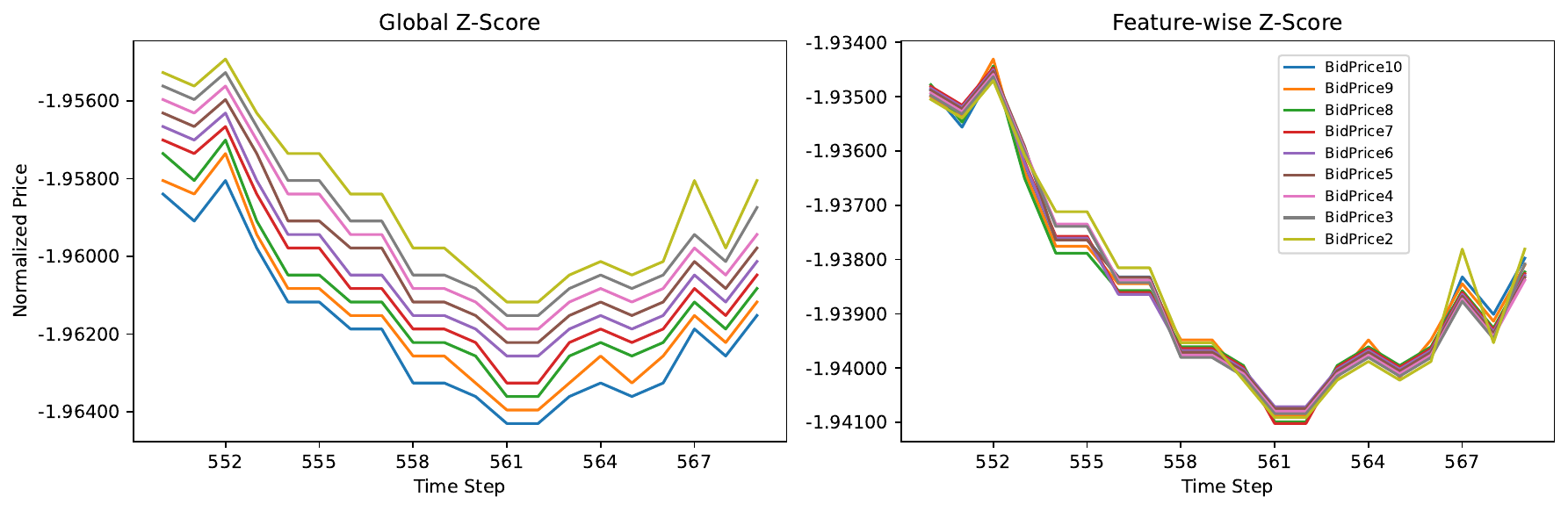
        }
        \caption{The commonly used feature-wise
        z-score normalization breaks the price-level constraint of LOB, while the Global Z-score does not.}
        \label{fig:compare_methods1}
    \end{figure}

\begin{figure}[htpb]
        \centering
        \includegraphics[width=0.49\textwidth]{
            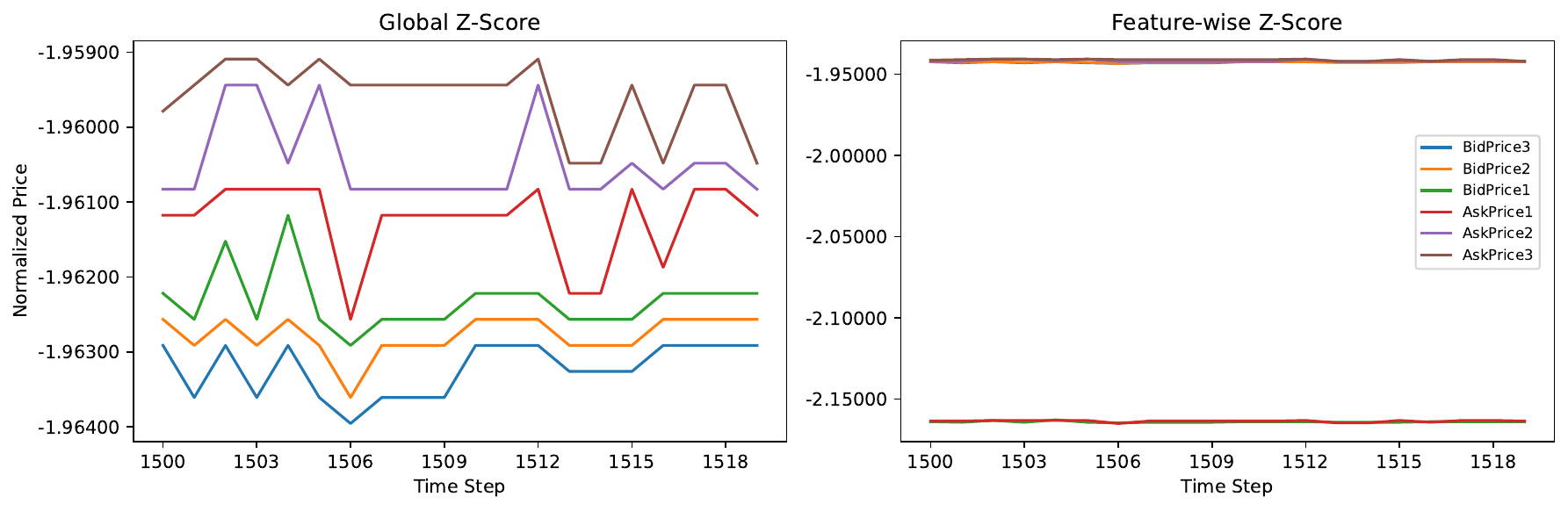
        }
        \caption{The commonly-used feature-wise z-score normalization introduced large deviations among different levels, while the Global Z-score does not.}
        \label{fig:compare_methods2}
    \end{figure}

    Our study shows that, although this method works effectively for many multivariate time series, it
    can unintentionally break essential LOB constraints in two notable ways. First,
    it disrupts the relative ordering among different price levels (e.g., $a_{i}^{p}
    < a_{j}^{p}$ for $i < j$ on the ask side), as illustrated in Fig.~\ref{fig:compare_methods1}.
    Second, it can introduce disproportionately large deviations between the best
    bid/ask price and other price levels, as shown in Fig.~\ref{fig:compare_methods2}.
    Both issues reduce the model’s ability to capture the essential structure of LOB data,
    particularly in tasks such as reconstruction, where preserving a consistent price
    hierarchy is crucial.

    To better preserve the internal data constraints of LOB, an alternative is to normalize price and
    volume collectively. Specifically, we can write: 
    \begin{equation}
        p_{ij}' = \frac{p_{ij}- \mu_{\text{price}}}{\sigma_{\text{price}}},
    \end{equation}
    \begin{equation}
        v_{ij}' = \frac{v_{ij}- \mu_{\text{volume}}}{\sigma_{\text{volume}}}, \label{eq:group_based_normalization}
    \end{equation}
    \noindent where $\mu_{\text{price}}, \sigma_{\text{price}}, \mu_{\text{volume}}, \sigma_{\text{volume}}$ denote the mean and standard deviation of all price and volume entries within the same LOB snapshot. Prices (both bid and ask) are normalized using $\mu_{\text{price}}$ and $\sigma_{\text{price}}$, and volumes using $\mu_{\text{volume}}$ and $\sigma_{\text{volume}}$. 
    This global z-score normalization ensures that the distinct relationships among
    levels remain more intact, potentially improving a model’s ability to reconstruct
    or learn from structural patterns in LOB data.

    In the experiments that follow, we will compare these normalization
    strategies across both reconstruction-based and downstream predictive tasks.
    By examining performance under feature-wise vs.\ global z-score normalization, we
    can ascertain whether preserving the inherent constraints between price features
    leads to better modeling results or more robust predictive power.

    \subsubsection{Segmenting}
    After normalizing the LOB data, we partition it into training and test
    subsets in an 4:1 ratio. Specifically, we split each full sequence (from 9:30 to 14:57) of LOB snapshots (in total 4740 snapshots a stock a day) so that the earliest 80\% of the data is used for training,
    while the remaining 20\% is reserved for testing. Subsequently, we employ a sliding-window
    method to segment each portion into contiguous blocks of 100 snapshots with the window moving step-size of 1 snapshot. More
    formally, each window is represented as a three-dimensional tensor $\mathbf{S}
    _{t}\in \mathbb{R}^{100 \times 40 }$, where $\mathbf{S}_{t}$ corresponds to
    10 price levels in 4 data fields (bid price, bid volume, ask price, ask
    volume), or 40 LOB features, and 100 consecutive time steps.

    This segmentation process not only respects temporal ordering but also facilitates
    batch-based processing for model training and evaluation. A window size of
    100 snapshots is commonly adopted in LOB-related studies, providing a well-established
    balance between context and complexity. At our 3-second sampling rate, each 100-snapshot
    window corresponds to roughly five minutes of real market activity—an interval
    sufficiently long to capture intraday patterns yet short enough to maintain computational
    tractability.

    \subsubsection{Labeling}
    To assign labels for downstream tasks, we compute the price trend based on
    the averaged mid-price of subsequent LOB snapshots. The mid-price at time step $t$ is defined as:
    \begin{equation}
        m_t = \{\frac{a_1^p+b_1^p}{2}\}_t,\label{eq:mid_price}
    \end{equation}
    \noindent where $a_1^p$ and $b_1^p$ denote the best ask and bid price at time $t$, respectively. Specifically, for the
    last time step (i.e., the $100$-th snapshot) in a given segment, the trend is determined based on the averaged
    mid-price over the following 5 snapshots and threshold using a predefined $\delta
    = 0.001$, as suggested in \cite{balch2019evaluatetradingstrategiessingle}. The trend labels are calculated with Eq.(\ref{eq:trend_prediction}).
    This labeling method ensures that the dataset captures meaningful market
    trends and provides well-defined targets for model training and evaluation.

    \begin{figure}[!ht]
        \centering
        \includegraphics[width=0.35\textwidth,]{
            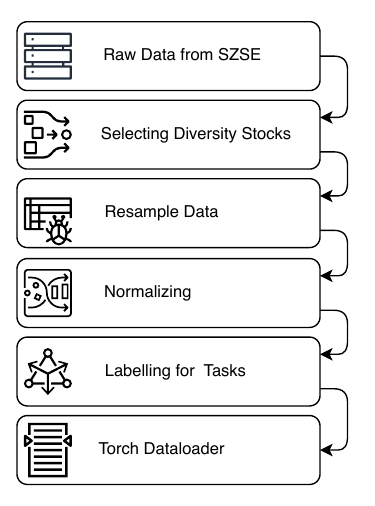
        }
        \caption{The Data Preprocessing Pipeline}
        \label{fig:data_preprocessing}
    \end{figure}

    \subsubsection{Dataloader Construction}

    After preprocessing and labeling, the data is converted into PyTorch \texttt{Dataset} and wrapped with a \texttt{DataLoader}\footnote{https://pytorch.org/docs/stable/data.html} for efficient batch training. This step enables streamlined data feeding during model training and evaluation, supporting shuffling, mini-batching, and parallel loading.

    By following this approach, the constructed dataset reflects the dynamic
    characteristics of the SZSE market while maintaining robustness and
    consistency for model evaluation. The full data preprocessing pipeline is illustrated
    in Fig.~\ref{fig:data_preprocessing}.

    \subsection{Baseline Models}
    \label{sec:model_selecting} In line with recent advances in time-series analysis
    and deep learning, we have consolidated 9 neural network models into our
    benchmark framework. They are divided into three conceptual groups based on their
    architecture designs and the degree to which they incorporate domain-specific
    considerations for LOB, more details are provided in the Tab.~\ref{tab:selected_models}.

    \begin{itemize}
        \item \textbf{Foundational Baselines:} This group comprises classic
            neural networks widely used across diverse tasks. CNNs\cite{tsantekidis2020cnn2} excel at capturing local
            spatial patterns within LOB snapshots, while LSTM\cite{tsantekidis2017lstm}
            networks and Transformers\cite{zerveas2021transformer} can handle temporal dependencies . We
            adapt these 'pure' models that have not been explicitly designed for
            LOB. Evaluating their performance establishes a comprehensive foundation
            for subsequent, more specialized approaches.

        \item \textbf{Generic Time-Series Models:} The second group encompasses
            several advanced architectures (iTransformer\cite{liu2024itransformer},
            TimesNet\cite{wu2022timesnet}, TimeMixer\cite{wang2024timemixer}) that
            have demonstrated state-of-the-art performance on general time-series
            tasks (e.g., forecasting, anomaly detection) in various benchmark
            studies. These models, sourced from the top-ranked entries in the
            latest Awesome Timeseries Library\footnote{\url{https://github.com/thuml/Time-Series-Library}},
            are included in our framework to reveal the trade-off between broad
            applicability and domain specificity. Despite their robust performance
            on large, heterogeneous MTS datasets, these architectures have not been expressly
            designed for LOB forecasting.

        \item \textbf{LOB-Specific Methods:} For the last group, our benchmark incorporates
            three specialized methods developed to address specific challenges in
            LOB tasks. While these methods ( i.e., DeepLOB\cite{zhang2019deeplob},
            TransLOB\cite{wallbridge2020translob}, and SimLOB\cite{liyuanzhe2024simlob})
            are tailored to address LOB-specific problems, only SimLOB explicitly
            focuses on learning a separate representation of LOB data. In
            contrast, the other two adopt the end-to-end frameworks optimized for specific downstream
            LOB tasks.
    \end{itemize}

        \small
    \begin{table}[!t]
        \centering
        \caption{Selected Models}
        \label{tab:selected_models}
        \setlength{\tabcolsep}{3pt} % 调整列间距（可选）
        \begin{tabular}{|c|c|c|c|c|}
            \hline
            \textbf{Group} & \textbf{Name}  & \textbf{Params}(million) & \textbf{Size}(MB) \\
            \hline
            \hline
            \multirow{3}{*}{Baselines} 
            & CNN2\cite{tsantekidis2020cnn2}               & 16.06 & 61.27 \\
            \cline{2-4}
            & LSTM\cite{tsantekidis2017lstm}              & 15.44 & 58.88 \\
            \cline{2-4}
            & Transformer\cite{zerveas2021transformer}  & 35.51 & 135.47 \\
            \hline
            \hline
            \multirow{3}{*}{Generic}
            & iTransformer\cite{liu2024itransformer}  & 20.54 & 78.35 \\
            \cline{2-4}
            & TimesNet\cite{wu2022timesnet}      & 21.68 & 82.72 \\
            \cline{2-4}
            & TimeMixer\cite{wang2024timemixer}     & 16.59 & 63.29 \\
            \hline
            \hline
            \multirow{3}{*}{LOB-Specific}
            & DeepLOB\cite{zhang2019deeplob}               & 15.42 & 58.84 \\
            \cline{2-4}
            & TransLOB\cite{wallbridge2020translob}      & 49.44 & 188.60 \\
            \cline{2-4}
            & SimLOB\cite{li2024simlob}       & 24.66 & 94.06 \\
            \hline
        \end{tabular}
    \end{table}
    \normalsize

    \subsection{Downstream Tasks}

    LOB data underpins a variety of applications in financial markets, enabling
    tasks such as price prediction, anomaly detection, and market behavior
    modeling. These tasks are crucial for understanding and capitalizing on the ever-changing
    market dynamics. In this work, we focus on three representative downstream tasks—price
    trend prediction, imputation, and reconstruction. Together, these tasks comprehensively assess how well a representation model captures both the temporal
    patterns and structural constraints of LOB.

    \paragraph{Price Trend Prediction}
    \label{pra:price_trend} This classification task targets the future
    direction of mid-price movement (e.g., upward, downward, or neutral) based
    on a given LOB snapshot at time $t$, as defined by Eq.~\ref{eq:mid_price}. 

    To determine the future trend, the model predicts a label $y_{t}$ as follows:
    \begin{equation}
        \label{eq:trend_prediction}y_{t}=
        \begin{cases}
            1,  & \text{if }\bar{m}_{t:t+\Delta t}> (1 + \delta) m_{t}, \\
            -1, & \text{if }\bar{m}_{t:t+\Delta t}< (1 - \delta) m_{t}, \\
            0,  & \text{otherwise}.
        \end{cases}
    \end{equation}
    \noindent where $\bar{m}_{t:t+\Delta t}$ denotes the averaged mid-price over
    the interval $( t, t+\Delta t]$, $\Delta_{t}$ is the prediction horizon, and
    $\delta$ specifies the minimum relative change required to classify a movement
    as significant. A label of $1$ indicates a significant upward trend, $-1$ a significant downward trend, and $0$ denotes no significant change.

    \paragraph{LOB Imputation}
    This regression task focuses on filling in missing LOB snapshots to ensure
    temporal consistency and structural validity. Given a sequence of LOB snapshots
    $\mathbf{S}_{t}$, the model aims to reconstruct the missing snapshot
    $\mathbf{L}_{mask_i},i\in \{1,2,\cdots,k\}$ for $mask_{i}\in [1, T]$. The objective
    is to minimize: 
    \begin{equation}
        \mathcal{L}_{\text{impu}}= \frac{1}{k}\sum_{i=1}^{k}\left\lVert \mathbf{L}
        _{mask_i}- \hat{\mathbf{L}}_{mask_i}\right\rVert^{2},
    \end{equation}
    
    \noindent where $\hat{\mathbf{L}}_{mask_i}$ is the $mask_i$-th snapshot in the true LOB series $\mathbf{S}_t$,
    ${\mathbf{L}}_{mask_i}$ is the corresponding prediction.

    \paragraph{LOB Reconstruction}
    Another fundamental approach to representation learning is reconstruction of
    the full LOB from its latent encoding. Here, a model learns to compress a
    LOB series $\mathbf{S}_{t}$ into a lower-dimensional representation and then
    reconstruct it as $\hat{\mathbf{S}}_{t}$. The reconstruction loss is typically
    measured using metrics such as the mean square error(MSE) or a weighted MSE that accounts for the relative
    importance of different LOB features. By quantifying the ability to
    accurately rebuild the original LOB data, this task serves as an unsupervised
    benchmark for evaluating how well the representation captures key structural
    and temporal characteristics of LOB data.

    \subsection{Metrics}
    \label{sec:evaluation_protocols}

    To thoroughly assess the quality of learned LOB representations, we adopt
    the above-mentioned three commonly studied time series tasks: \emph{reconstruction}, \emph{prediction},
    and \emph{imputation}. Each task targets a different aspect of modeling performance,
    providing a multifaceted view of how well candidate models capture and
    exploit LOB-specific dynamics.

    \paragraph{Reconstruction}
    This task measures a model's ability to accurately reconstruct raw LOB inputs
    from its learned latent representations. Let
    $\mathbf{S_t}\in \mathbb{R}^{100 \times 40}$ be the real LOB series data starting at
    time $t$ and $\hat{\mathbf{S_t}}$ be the model's reconstruction. We compare models
    using MSE, mean absolute error (MAE), and weighted MSE (wMSE),
    given by: 
    \begin{equation}
        \label{eq:mse}\text{MSE}= \frac{1}{100\times 40}\sum_{i=1}^{100}\sum_{j=1}
        ^{40}(x_{ij}- \hat{x}_{ij})^{2},
    \end{equation}
    \begin{equation}
        \label{eq:mae}\text{MAE}= \frac{1}{100\times 40}\sum_{i=1}^{100}\sum_{j=1}
        ^{40}\big| x_{ij}- \hat{x}_{ij}\big|,
    \end{equation}
    \begin{equation}
        \label{eq:wmse}\text{wMSE}= \frac{1}{W}\sum_{j=1}^{40}w_{j}\left[ \sum_{i=1}
        ^{100}(x_{ij}- \hat{x}_{ij})^{2}\right],
    \end{equation}
    \noindent where $w_{ij}$ is a weight factor indicating the relative importance
    of each price/volume entry, and $W = \sum_{j=1}^{40}w_{j}$. Decomposing these
    metrics by prices and volumes further reveals whether certain architectures focus
    more on crucial features (e.g., best bid/ask prices) than others.

    Since prices and volumes in LOB data carry fundamentally different meanings, we further introduce two specialized metrics to separately evaluate model performance on these two components: $\mathcal{L}_{\text{price}}$ and $\mathcal{L}_{\text{volume}}$. This decomposition allows us to better understand how well a model captures structural differences within the LOB.

    \begin{equation}
        \mathcal{L}_{\text{price}} = \frac{1}{100 \times d_p} \sum_{i=1}^{100} \sum_{j=1}^{d_p} (x_{ij} - \hat{x}_{ij})^2,
    \end{equation}
    \begin{equation}
        \mathcal{L}_{\text{volume}} = \frac{1}{100 \times d_v} \sum_{i=1}^{100} \sum_{j=d_p+1}^{40} (x_{ij} - \hat{x}_{ij})^2,
    \end{equation}
    
    \noindent where $d_p$ and $d_v$ represent the number of price and volume features per snapshot respectively. In our setting, $d_p = 20$ and $d_v = 20$.

    In this paper, we argue that the above standard metrics are less effective, and propose a novel composite loss
    function, called $\mathcal{L}_{\text{All}}$, to more holistically evaluate
    reconstruction quality while incorporating LOB-specific priors. It is defined
    as: 
    \begin{equation}
        \label{eq:allloss}\mathcal{L}_{\text{All}}= \alpha\text{MSE}+ (1-\alpha)\text{wMSE}
        + \lambda \cdot \mathcal{L}_{\text{reg}},
    \end{equation}
    \begin{equation}
        \label{eq:regloss}\mathcal{L}_{\text{reg}}=\frac{1}{20-1}\sum_{p=2}^{20}\text{ReLU}
        (price_{p-1}-price_{p})
    \end{equation}
    \noindent where $\mathcal{L}_{\text{reg}}$ is a structural regularization
    term penalizing violations of price ordering (e.g., inverted bid/ask levels);
    $\alpha$ and $\lambda$ are tunable coefficients controlling the strength of \text{wMSE}
    and penalization; $p rice_{i}$ denotes the $i$-th lowest price in the LOB
    cross bid and ask sides; \text{ReLU} is the commonly used rectified linear unit
    function in DL.

    This proposed loss encourages reconstructions that are not only numerically accurate but
    also structurally consistent with real-world LOB dynamics. By combining
    point-wise and structure-aware terms, $\mathcal{L}_{\text{All}}$ guides models
    to better preserve economically meaningful patterns, such as best bid/ask
    stability and monotonicity across price levels.

    \paragraph{Prediction}
    In this classification task, the goal is to predict the future direction of the
    mid-price based on its learned representations. Given a segment
    $\mathbf{S}_{t}$, the models classify whether the average mid-price in a subsequent
    time window rises, falls, or steady. We use cross-entropy (CE) as the
    evaluation metric:
    \begin{equation}
        \text{CE}= - \sum_{c=1}^{C}y_{c}\log \hat{p}_{c},
    \end{equation}
    \noindent where $C$ is the number of trend classes (e.g., $\{+1, 0, -1\}$),
    $y_{c}$ is the true label for class $c$, and $\hat{p}_{c}$ is the predicted probability
    of class $c$. High accuracy and low CE indicate that the model captures
    relevant market cues for anticipating short-term price movements.

    \paragraph{Imputation}
    This regression-based task targets the model's capability to fill in missing
    data (i.e., masked time steps) within a given LOB sequence $\mathbf{S}_{t}$.
    Let $\mathcal{M}=\{L_{t+m_i}|i\in \left[ 1,k \right] \}$ be the set of
    masked entries in $\mathbf{S}_{t}$, and $\hat{\mathcal{M}}$ be the predicted
    LOB. We measure performance using MSE restricted to the masked region: 
    \begin{equation}
        \text{MSE}_{\mathcal{M}}= \frac{1}{k}\sum_{i\in \left[ 1,k \right] }( L_{t+m_i}
        - \hat{L}_{t+m_i})^{2}.
    \end{equation}
    \noindent A lower $\text{MSE}_{\mathcal{M}}$ indicates that the learned representation
    captures essential structural and temporal relationships, enabling more
    accurate recovery of unobserved prices and volumes.

    By jointly examining performance on reconstruction, prediction, and
    imputation, we obtain a comprehensive understanding of how well different
    architectures extract, retain, and utilize LOB-specific knowledge. This
    holistic view illuminates the trade-offs among various design choices, ultimately
    guiding the development of robust LOB representation learners.
    \label{sec:benchmark_workflow}

    \section{Experimental Studies}
    \label{sec:experiment}

    \begin{figure*}[tb]
        \centering
        \includegraphics[width=0.7\textwidth, keepaspectratio]{
            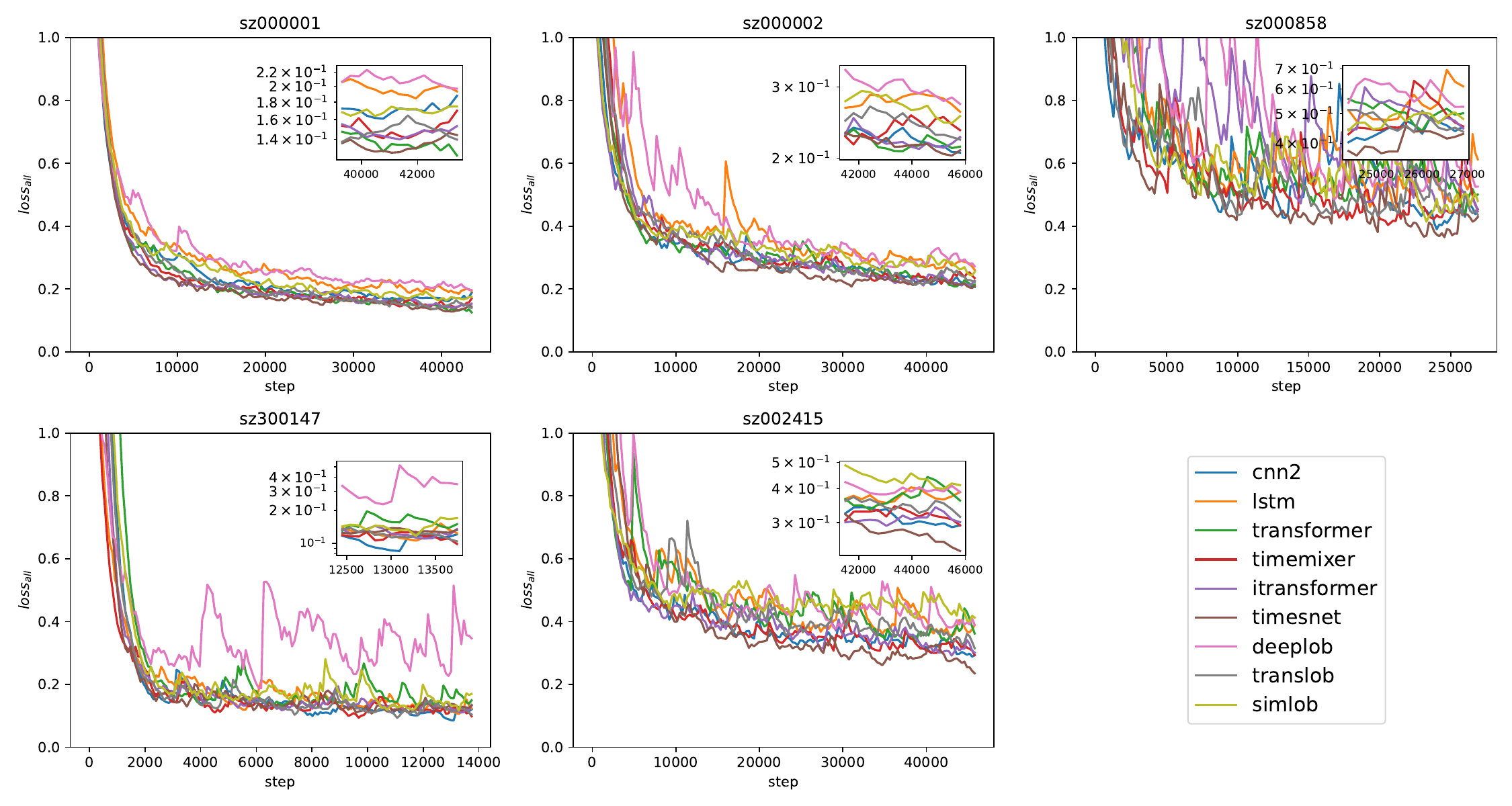
        }
        \caption{Training curves with the loss function
        $\mathcal{L}_{\text{All}}$ on the Reconstruction task across different
        models and different datasets.}
        \label{fig:result_exp1_reconstruction_loss}
    \end{figure*}

    To maintain consistency, we re-implemented each compared model using the \emph{PyTorch
    Lightning}\footnote{\url{https://lightning.ai}} framework, and tracked
    experimental runs via the \emph{Comet}\footnote{\url{https://www.comet.com}}
    library. Owing to network bandwidth constraints, we log training metrics
    every 50 batches rather than every batch. Furthermore, all models are
    configured so that their latent representation dimension is the same (i.e., 256 in all experiments). In order
    to accommodate diverse architectures, we introduce several fully connected
    layers before feeding the LOB data into encoder and after
    obtaining decoder output of all models, thereby ensuring correct input-output
    dimensions for every approach. Beyond these adjustments, model-specific
    parameters such as learning rates and layer sizes adhere to recommendations 
    from the original authors, striking a balance between faithful reproduction and
    framework-level consistency. We use the Adam optimizer for training all models, given its broad applicability across deep learning tasks and favorable convergence observed in our preliminary experiments. The number of training epochs is uniformly set to 100 to ensure fair comparison under consistent training conditions.
    
    All primary experiments are conducted on LOBench using a server equipped with two NVIDIA A6000
    GPUs and an Intel Xeon CPU (2.00\,GHz, 128 cores). This setup ensures consistent
    and fair comparisons of model execution times and resource usage across all benchmark
    approaches.

    \subsection{Group 1: Representation Tasks}

    \subsubsection{Experimental Setup}
    \label{sec:exp_sufficiency} In this experiment, we evaluate how effectively each
    model preserves the detailed structure of LOB data when mapping it into a common
    representation space. Specifically, we train all representation-learning models
    to encode the data into identical dimensional embeddings, and then employ a
    uniform decoder to reconstruct the original LOB inputs. Our training procedure
    leverages $\mathcal{L}_{\text{All}}$ as Eq.(\ref{eq:allloss}). We conduct
    these reconstruction experiments on the selected five diverse China A-share
    datasets to ensure robustness in performance assessment.

    Multiple metrics are tracked on both the training and test sets, including MSE,
    wMSE and MAE, along with decomposed variants for price and volume. Training time
    is also monitored to assess computational efficiency. Comparisons of these
    metrics across different models reveal the extent to which each
    representation preserves fine-grained LOB characteristics, and address the second
    research question regarding the sufficiency of the learned representations.

    \subsubsection{Results}
    \small
    \begin{table}[tb]
        \centering
        \caption{Models Training time in Reconstruction Experiments(in seconds).}
        \label{tab:result_exp1_time} \resizebox{\columnwidth}{!}{%
        \begin{tabular}{|l|r|r|r|r|r|}
            \hline
            \textbf{Model}        & \textbf{sz000001} & \textbf{sz000002} & \textbf{sz000858} & \textbf{sz300147} & \textbf{sz002415} \\
            \hline
            \textbf{cnn2}         & 2047     & 3159     & 1404     & 1508     & 2257     \\
            \hline
            \textbf{lstm}         & 2553     & 3206     & 1376     & 1368     & 2261     \\
            \hline
            \textbf{transformer}  & 5498     & 8131     & 3412     & 3236     & 5646     \\
            \hline
            \textbf{timemixer}    & 2675     & 4149     & 1739     & 1626     & 2914     \\
            \hline
            \textbf{itransformer} & 2979     & 3602     & 1585     & 1603     & 2559     \\
            \hline
            \textbf{timesnet}     & 9905     & 15399    & 6189     & 6099     & 10354    \\
            \hline
            \textbf{deeplob}      & 3557     & 4515     & 1838     & 1918     & 3095     \\
            \hline
            \textbf{translob}     & 3428     & 5160     & 2245     & 2095     & 3531     \\
            \hline
            \textbf{simlob}       & 4756     & 6677     & 2731     & 2568     & 4592     \\
            \hline
        \end{tabular}
        }
    \end{table}
    \normalsize

    Firstly, an examination of the reconstruction training curves (Fig.~\ref{fig:result_exp1_reconstruction_loss})
    reveals that with the exception of LSTM and DeepLOB, the remaining models
    show no significant differences during training. Notably, the overall
    quality of the learned representations is correlated with the dataset itself.
    However, models incorporating Transformer architectures consistently
    outperform those based on other architectures. In particular, TimesNet
    achieves superior performance on both the training and validation sets(see
    Tab.~\ref{tab:result_exp1_mse_wmse}). It is also worth noting that, after
    incorporating a weighted loss term into the training objective, nearly all models
    exhibit lower weighted MSE compared to the standard MSE. This improvement
    confirms that emphasizing critical features during training leads to more
    accurate reconstructions.

\begin{figure}[tb]
        \centering
        \includegraphics[width=0.5\textwidth, keepaspectratio]{
            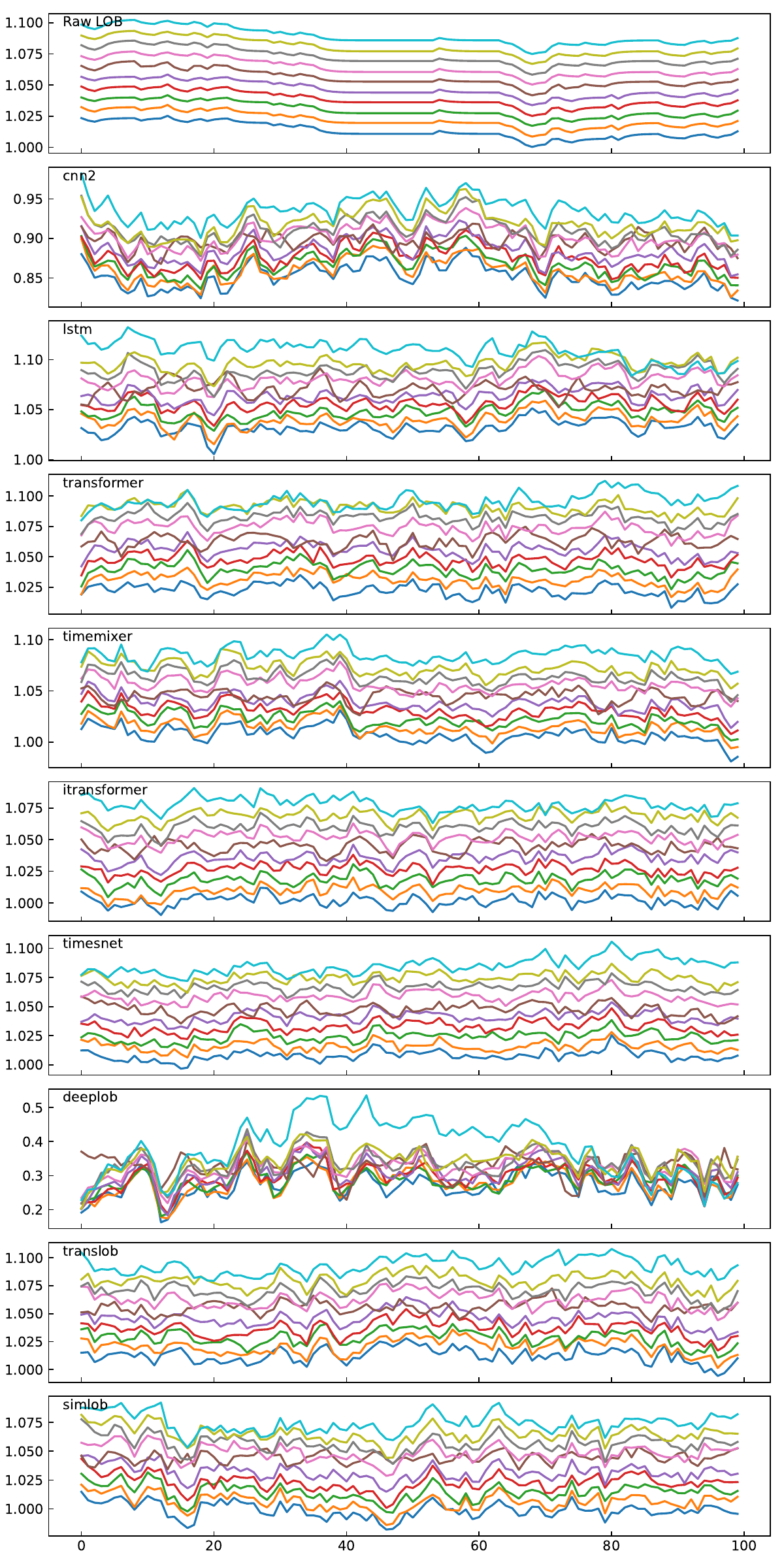
        }
        \caption{Example of reconstructed LOB data by different models.}
        \label{fig:result_exp1_reconstruction_example}
    \end{figure}

    Secondly, by randomly selecting a data segment and reconstructing the LOB across
    nine different models, the visualized price data (Fig.~\ref{fig:result_exp1_reconstruction_example})
    further supports these findings. The reconstructions produced by DeepLOB and
    CNN display significant shifts, failing to clearly separate individual price
    curves. This suggests that relying solely on convolutional architectures may
    not suffice for the intricate task of LOB representation. In contrast, the
    other models manage to reconstruct the original LOB curve more faithfully. These
    reconstruction have minimal occurrences of price level crossovers, which is
    likely a benefit of the penalty term integrated into the loss function.

    Additionally, Tab.~\ref{tab:result_exp1_time} presents the training time required
    for each model. Although TimesNet exhibits the best overall performance, its
    recurrent structure results in longer training times. Despite having the
    highest number of trainable parameters, the Transformer and TransLOB models do
    not achieve outstanding performance. This indicates that, compared to sheer model
    complexity, choosing an appropriate architecture is more beneficial for
    effective LOB data representation.

    Finally, these experiments collectively demonstrate that, once LOB data is
    mapped to a low-dimensional space, most models are capable of reconstructing
    the original space while preserving the majority of its intrinsic features.
    The low MSE and weighted MSE losses (on the order of 1e-3) observed on the
    validation set further indicate the sufficiency of the learned
    representations. In essence, by selecting an appropriate representation
    model, LOB data can be effectively encoded into a low-dimensional form that remains
    both informative and faithful to the original structure. Thus addresses the
    first fundamental issues-the representation of LOB data is sufficient for
    downstream tasks, as it does not lose the essential information.

    More detailed numerical results on the test sets for each model, along with further
    quantitative analysis, can be found in Tab.~\ref{tab:result_exp1_mse_wmse}-\ref{tab:result_exp1_pv}.
    More comparative results upon additional metrics (e.g., MSE, MAE, wMSE, $\mathcal{L}
    _{price}$, $\mathcal{L}_{volume}$, $\mathcal{L}_{reg}$) are available in
    supplementary materials.

    \small
    \begin{table}[tb]
        \centering
        \caption{Reconstruction error metrics on valid datasets.}
        \label{tab:result_exp1_mse_wmse} \resizebox{\columnwidth}{!}{%
        \begin{tabular}{|l|l|r|r|r|r|r|}
            \hline
            \textbf{Model}                         & Metrics & sz000001 & sz000002 & sz000858 & sz300147 & zs002415 \\
            \hline
    
            \multirow{3}{*}{\textbf{cnn2}} & MSE      & 0.0921 & \textbf{0.0778} & 0.3878 & 0.0073 & 0.2863 \\
            \cline{2-7}                   & wMSE     & 0.0222 & 0.0272 & 0.1951 & 0.0075 & 0.0462 \\
            \cline{2-7}                  & MAE      & 0.1320 & \textbf{0.1116} & 0.1573 & 0.0400 & 0.1806 \\
            \hline
    
            \multirow{3}{*}{\textbf{lstm}} & MSE      & 0.1417 & 0.1275 & 0.2502 & 0.0117 & 0.3446 \\
            \cline{2-7}                   & wMSE     & 0.0453 & 0.0350 & 0.1763 & 0.0146 & 0.0534 \\
            \cline{2-7}                  & MAE      & 0.1519 & 0.1408 & \textbf{0.1343} & 0.0454 & 0.1788 \\
            \hline
    
            \multirow{3}{*}{\textbf{transformer}} & MSE      & \textbf{0.0531} & 0.1286 & \textbf{0.2197} & 0.0109 & 0.1827 \\
            \cline{2-7}                   & wMSE     & \textbf{0.0146} & 0.0740 & 0.1058 & 0.0152 & 0.0555 \\
            \cline{2-7}                  & MAE      & \textbf{0.0995} & 0.1457 & 0.1686 & 0.0418 & \textbf{0.1622} \\
            \hline
    
            \multirow{3}{*}{\textbf{timemixer}} & MSE      & 0.0651 & 0.1225 & 0.2633 & 0.0105 & \textbf{0.1271} \\
            \cline{2-7}                   & wMSE     & 0.0211 & 0.1259 & \textbf{0.0820} & 0.0179 & \textbf{0.0345} \\
            \cline{2-7}                  & MAE      & 0.1191 & 0.1326 & 0.1701 & \textbf{0.0344} & 0.1453 \\
            \hline
    
            \multirow{3}{*}{\textbf{itransformer}} & MSE      & 0.0625 & 0.1551 & 0.2343 & \textbf{0.0045} & 0.1329 \\
            \cline{2-7}                   & wMSE     & 0.0184 & \textbf{0.0452} & 0.1205 & \textbf{0.0044} & 0.0438 \\
            \cline{2-7}                  & MAE      & 0.1102 & 0.1406 & 0.1541 & \textbf{0.0317} & 0.1477 \\
            \hline
    
            \multirow{3}{*}{\textbf{timesnet}} & MSE      & 0.0983 & 0.0852 & \textbf{0.2084} & 0.0064 & 0.1348 \\
            \cline{2-7}                   & wMSE     & 0.0186 & \textbf{0.0246} & \textbf{0.0692} & 0.0079 & 0.0467 \\
            \cline{2-7}                  & MAE      & 0.1145 & 0.1222 & \textbf{0.1315} & 0.0347 & \textbf{0.1375} \\
            \hline
    
            \multirow{3}{*}{\textbf{deeplob}} & MSE      & 0.2063 & 0.1144 & 0.2676 & 0.0075 & 0.1436 \\
            \cline{2-7}                   & wMSE     & 0.0388 & 0.0596 & 0.1042 & 0.0075 & 0.0460 \\
            \cline{2-7}                  & MAE      & 0.1665 & 0.1365 & 0.1565 & 0.0425 & 0.1803 \\
            \hline
    
            \multirow{3}{*}{\textbf{translob}} & MSE      & 0.0756 & 0.0930 & 0.3399 & 0.0173 & 0.2654 \\
            \cline{2-7}                   & wMSE     & 0.0310 & 0.0308 & 0.0783 & 0.0227 & 0.0587 \\
            \cline{2-7}                  & MAE      & 0.1129 & 0.1285 & 0.1577 & 0.0476 & 0.1721 \\
            \hline
    
            \multirow{3}{*}{\textbf{simlob}} & MSE      & 0.0798 & 0.3467 & 0.1973 & \textbf{0.0037} & 0.2097 \\
            \cline{2-7}                   & wMSE     & 0.0212 & 0.1073 & 0.0962 & \textbf{0.0014} & 0.0632 \\
            \cline{2-7}                  & MAE      & 0.1245 & 0.2072 & 0.1371 & 0.0348 & 0.1682 \\
            \hline
     
    \end{tabular}
    }
    \end{table}
    \normalsize

    \subsection{Group 2: Downstream Tasks}
    \label{sec:exp_downstream}
    \begin{figure}[tb]
        \centering
        \includegraphics[width=0.80\textwidth, keepaspectratio]{
            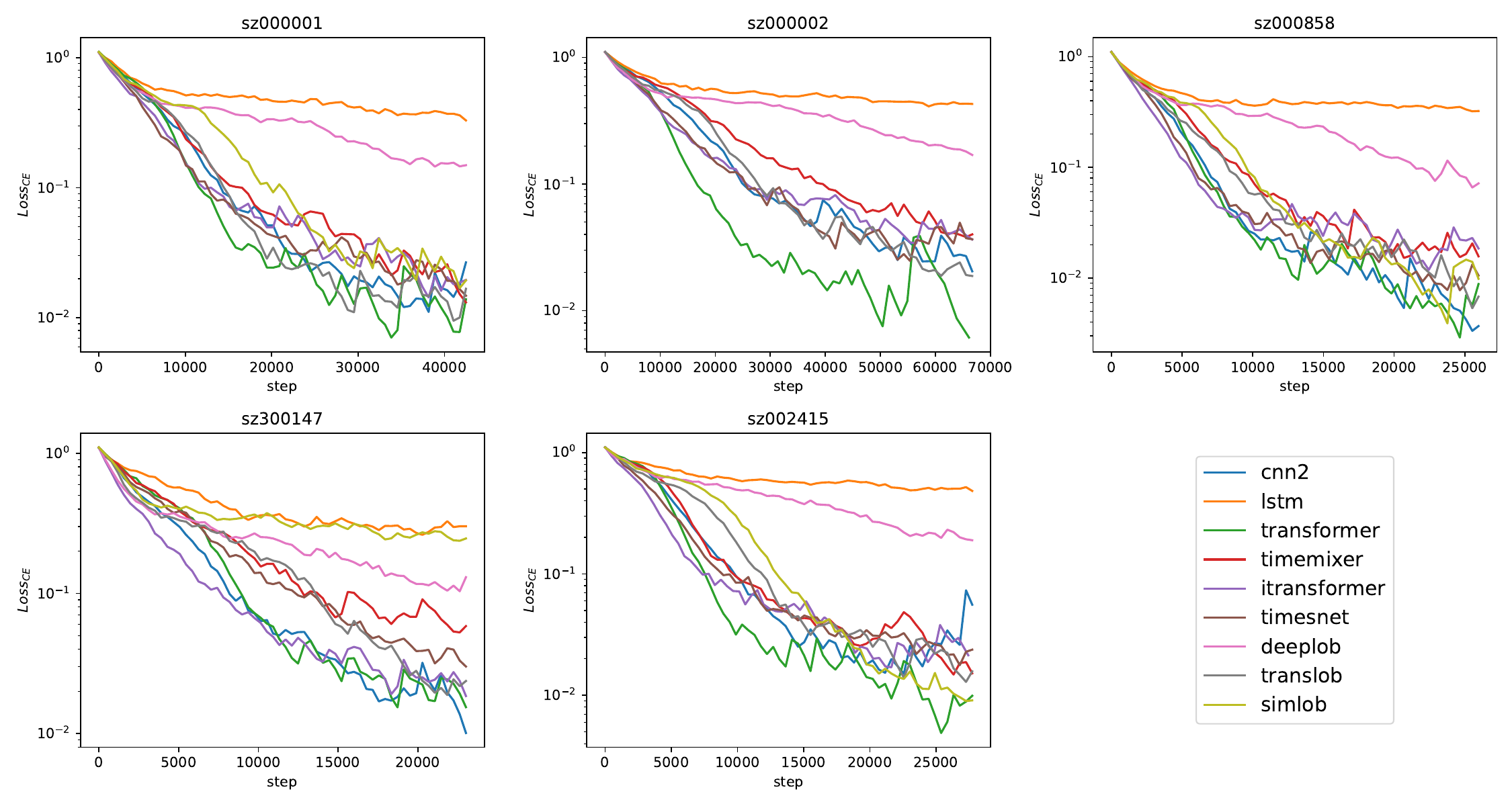
        }
        \caption{Training curves on the Prediction task.}
        \label{fig:result_exp2_prediction}
    \end{figure}
    \begin{figure}[h]
        \centering
        \includegraphics[width=0.80\textwidth, keepaspectratio]{
            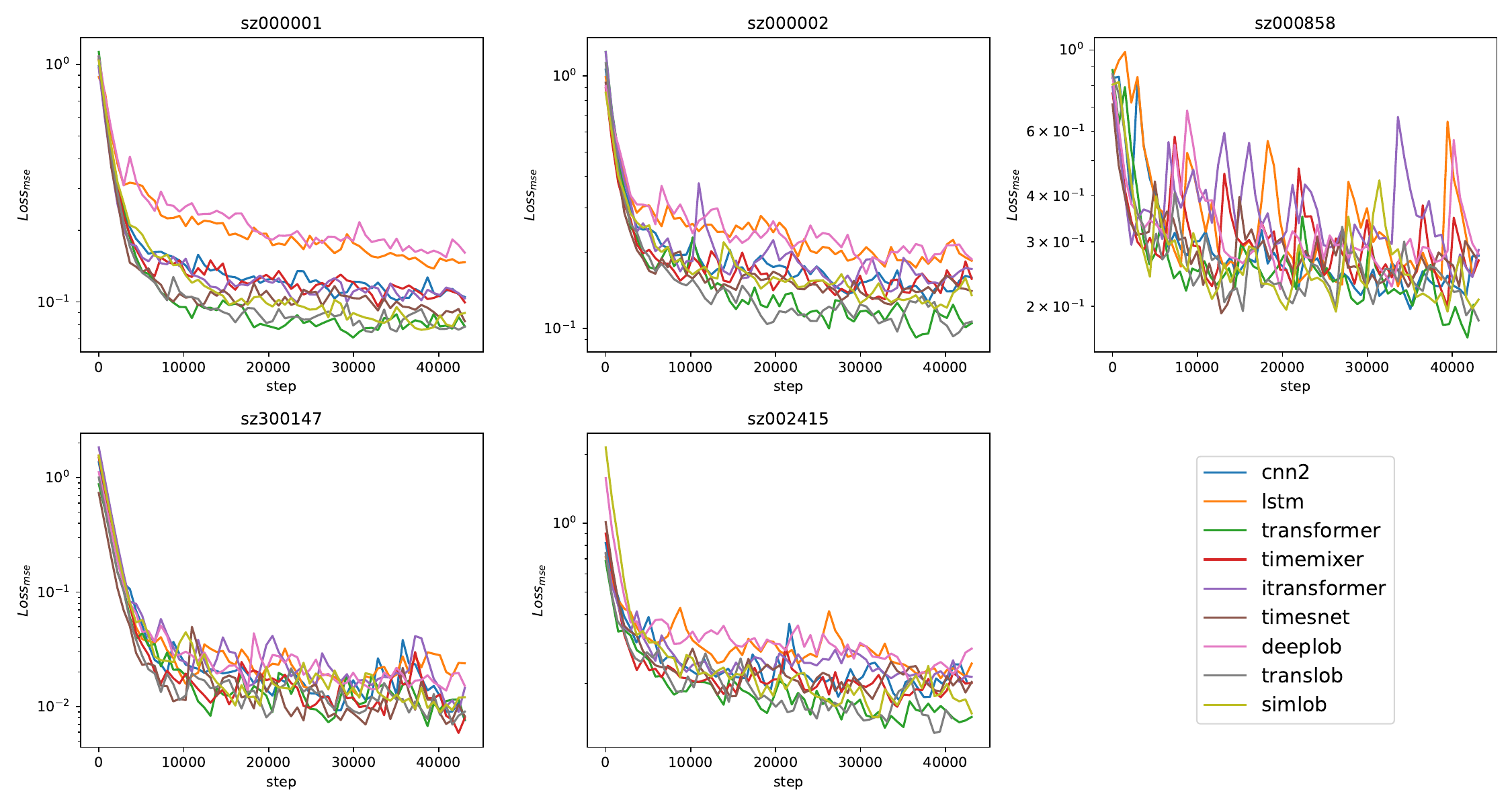
        }
        \caption{Training curves on the Imputation task.}
        \label{fig:result_exp2_imputation}
    \end{figure}

    \subsubsection{Experimental Setup}

    In this set of experiments, we aim to assess whether the representations
    learned through our proposed framework can effectively enhance performance
    on downstream tasks. Specifically, we replace the original decoder of the
    representation model with a simple multi-layer fully connected network that
    maps the learned representations to the desired output. By leveraging the universal
    representation model (i.e., the encoder from Group 1) to transform the LOB data
    into a rich, low-dimensional representation, a straightforward method is sufficient
    to address a variety of downstream tasks. This group of experiments
    primarily to address the second fundamental issue-the necessity of a robust
    representation-by showing that better representation model can yield
    consistently better performance across different downstream tasks.

    We evaluate all models on two downstream tasks: trend prediction and data imputation.
    In the prediction task, the goal is to forecast the next price trend as defined
    in Eq.~\ref{eq:trend_prediction}, using $\Delta t = 5$ and $\delta = 0.0001$,
    following the setting in \cite{li2023deep}. And to make the class distribution more balanced,
    we do a down-sampling on the training set, where we randomly select the same number
    of samples from each class and rebuild the train dataset, validation dataset
    and test dataset. The imputation task aims to recover the missing values in the
    LOB data, in which we randomly mask 20\% of the LOB entries in the series $\mathbf{S_t}$
    during training, and train the model to recover the masked values. Since the
    prediction task is actually a classification problem, we adopt the cross-entropy loss.
    For the imputation task, we use the MSE loss. This simplicity is one of the key
    advantages we expect representation learning to bring to downstream tasks.

    % \begin{figure}[tb]
    %     \centering
    %     \includegraphics[width=0.48\textwidth, keepaspectratio]{
    %         figures/results/prediction_freeze_train_loss_curve.pdf
    %     }
    %     \caption{
    %         Comparison of End-to-end training and Representation Reuse on the Prediction task.
    %     }
    %     \label{fig:result_exp1_prediction_freeze}
    % \end{figure}

    \subsubsection{Results}
    In the prediction task (Fig.\ref{fig:result_exp2_prediction}), DeepLOB and LSTM
    consistently exhibit the poorest performance, while the remaining models
    maintain performance levels similar to those observed in the reconstruction
    task. Similarly, in the imputation task (Fig.\ref{fig:result_exp2_imputation}),
    DeepLOB and LSTM again under-perform relative to the other models. To more
    clearly illustrate the convergence behavior during training, we plotted the training
    curves using a log scale on the y-axis.

    Notably, on the sz300147 dataset, SimLOB performed poorly in the reconstruction
    task, which in turn adversely affected its performance in both the
    prediction and imputation tasks for that dataset. Conversely, on the
    sz000858 dataset, all models exhibited only average reconstruction performance,
    a trend that was subsequently reflected in their downstream task outcomes.

    In addition, we evaluated a frozen encoder approach using the pretrained encoder
    from Group 1, where only the downstream decoders were trained. Remarkably,
    this separated training strategy achieved performance equivalent to that of
    end-to-end training—where both the encoder and decoder are trained
    simultaneously—albeit with a requirement for more training epochs. This result
    indirectly confirms the effectiveness of the LOB representation, as the
    capability of the models to solve downstream tasks is intrinsically linked
    to the quality of its learned representations. Moreover, by adopting this
    framework, researchers can concentrate on optimizing the representation model
    via reconstruction tasks, which do not require additional data labeling, thereby
    lowering the entry barrier for research. The resulting low-dimensional
    embeddings also facilitate the application of simpler and well-established
    machine learning methods for various downstream tasks, obviating the need for
    designing specialized architectures for each individual task.

    Overall, these results underscore the direct relationship between
    reconstruction quality—and by extension, the quality of the learned representations—and
    downstream task performance. Superior reconstruction capability is
    indicative of more effective representations, which consistently translate into
    enhanced performance in both prediction and imputation tasks for LOB data.

    \subsection{Experiment 3: Transferability of LOB Representations}
    \label{sec:exp_lob_specific}

    To further demonstrate the benefits of LOB representation learning, we
    conducted a supplementary experiment to assess model transferability.

    \subsubsection{Experiment Setup}
    Given that the structural characteristics of LOB data are largely consistent
    across different datasets—even though trends and dynamics may vary—the
    learned representations should exhibit high transferability. In this experiment,
    we first trained a prediction model on the SZ000001 dataset and directly
    evaluated its performance on other datasets, recording the corresponding accuracy
    on their test sets. In addition, we selected a relatively under-performing model
    on SZ000001 (SimLOB) and froze its encoder weights. We then fine-tuned only
    the downstream decoder on the target datasets using a limited dataset (100 batches,
    approximately 20\% of the total training data) and compared the results with
    those obtained from the directly transferred model.
    \small 
    \begin{table*}[htbp]
        \centering
        \caption{Transfer performance of models trained on sz000001 and tested on other datasets.}
        \label{tab:result_transfer} 
        \begin{tabular}{|l|c|c|c|c|c|c|c|c|c|c|}
            \hline
             \textbf{Dataset}& \multicolumn{2}{c|}{sz000002} & \multicolumn{2}{c|}{sz000858} & \multicolumn{2}{c|}{sz300147} & \multicolumn{2}{c|}{sz002415} & Mean & Mean \\
            \hline
            \textbf{Model}& Precision & Recall & Precision & Recall & Precision & Recall & Precision & Recall & Recall & Precision \\
            \hline
            cnn2 & 0.6206 & 0.6114 & 0.6174 & 0.5931 & 0.7001 & 0.6497 & 0.5774 & 0.5649 & 0.6048 & 0.6289 \\
            \hline
            lstm & 0.7098 & 0.7108 & 0.6423 & 0.6227 & 0.6890 & 0.6623 & 0.6490 & 0.6487 & 0.6611 & 0.6725 \\
            \hline
            transformer & 0.6219 & 0.6239 & 0.6356 & 0.6424 & 0.7045 & 0.6243 & 0.5980 & 0.6002 & 0.6227 & 0.6400 \\
            \hline
            timemixer & 0.6360 & 0.6271 & 0.6256 & 0.5845 & 0.6949 & 0.6424 & 0.6034 & 0.5839 & 0.6095 & 0.6399 \\
            \hline
            itransformer & 0.7231 & 0.7107 & 0.6728 & 0.6322 & 0.7502 & 0.6787 & 0.6674 & 0.6511 & 0.6682 & 0.7034 \\
            \hline
            timesnet & 0.6804 & 0.6744 & 0.6457 & 0.6242 & 0.7296 & 0.6572 & 0.6247 & 0.6153 & 0.6428 & 0.6701 \\
            \hline
            deeplob & 0.7110 & 0.7039 & 0.6783 & 0.6470 & \textbf{0.7922} & 0.7358 & 0.6497 & 0.6388 & 0.6814 & 0.7078 \\
            \hline
            translob & 0.6958 & 0.6911 & 0.6671 & 0.6515 & 0.7333 & 0.6386 & 0.6528 & 0.6460 & 0.6568 & 0.6873 \\
            \hline
            simlob & 0.6986 & 0.6971 & 0.6384 & 0.6066 & 0.7347 & 0.6559 & 0.6476 & 0.6437 & 0.6508 & 0.6798 \\
            \hline
            simlob-freeze & \textbf{0.7421} & \textbf{0.7548} & \textbf{0.7234} & \textbf{0.7392} & 0.7415 & \textbf{0.7406} & \textbf{0.6970} & \textbf{0.6983} & \textbf{0.7260} & \textbf{0.7332} \\
            \hline
        \end{tabular}
    \end{table*}
    \normalsize
    
    \subsubsection{Results}
    As shown in Tab.\ref{tab:result_transfer}, the fine-tuned model SimLOB-freeze
    outperformed other models that trained end-to-end. This result suggests that
    end-to-end models specifically designed for individual tasks have lower transferability.
    A plausible explanation is that, during the reconstruction task, the
    representation model is compelled to embed most of the key features of the
    original LOB data into a low-dimensional space at the latent layer. These features
    are generally invariant to the specific characteristics of the LOB, whereas
    end-to-end trained models may distribute the extraction of LOB features
    across various components of the network, thereby impeding effective transfer
    to other datasets.

    \section{conclusion}
    In this work, we have conducted the first systematic study of LOB representation learning. To address the lack of standardization in prior research, we introduced \textbf{LOBench}, a comprehensive benchmark providing curated datasets, unified preprocessing pipelines, and consistent evaluation metrics tailored to the unique characteristics of LOB data.

    Through extensive experiments, we systematically investigated three core aspects: (1) the importance of standardizing LOB data formats, preprocessing procedures, and downstream task evaluation protocols; (2) the sufficiency of learned representations in capturing key LOB properties through the reconstruction experiments; and (3) the necessity of adopting a representation learning paradigm over end-to-end modeling, in terms of improving generalization and model development efficiency.

    Our findings demonstrate that robust LOB representations not only faithfully preserve critical market information but also enable simpler downstream models to achieve competitive performance. By decoupling feature extraction from task-specific objectives, LOB representation learning provides a scalable and reusable solution for a wide range of financial modeling tasks.

    Looking forward, an important direction for future research is to investigate the geometric and semantic consistency of LOB representations—specifically, whether similar LOB states in the original space are mapped to proximate regions in the representation space. Establishing such consistency would enhance the interpretability of learned features and offer deeper insights into the underlying market dynamics. We plan to explore these aspects in subsequent studies.

    % \section*{Acknowledgment}

    % The preferred spelling of the word ``acknowledgment'' in America is without an
    % ``e'' after the ``g''. Avoid the stilted expression ``one of us (R. B. G.) thanks
    % $\ldots$''. Instead, try ``R. B. G. thanks$\ldots$''. Put sponsor
    % acknowledgments in the unnumbered footnote on the first page.

    %Bibliography
    \bibliographystyle{unsrt}  
    \bibliography{references}  

\end{document}